\documentclass[useAMS]{mn2e}
\usepackage{mnras_cite}
\usepackage{epsfig}

\def\H2{{\rm H}$_2$}

\begin{document}
\title[Disk Galaxy Evolution Along the Hubble Sequence]{Disk Galaxy Evolution Along the Hubble Sequence}\author[Marios Kampakoglou and  Joseph Silk]{Marios Kampakoglou and  Joseph Silk\\
Department of Physics, University of Oxford, Keble Road, Oxford OX1 3RH, United Kingdom (e-mail: mariosk,silk@astro.ox.ac.uk)\\
}

\maketitle

\begin{abstract}
Galaxy disks are characterised by star formation histories that vary
systematically along the Hubble sequence. We  study  global
star formation, incorporating supernova feedback, gas accretion
and enriched outflows in disks modelled by a  multiphase interstellar medium 
in a fixed gravitational potential. The star formation histories, gas
distributions and chemical evolution can be explained in a simple sequence of
models which are primarily regulated by the cold gas accretion history.
\end{abstract}

\section{Introduction}
The two most striking characteristics that define the Hubble sequence
are morphology and star formation activity. The former is best addressed by
numerical simulations \cite{abadi03,sharma05,robertson04,governato06}.
However the latter aspects are so complex that most discussions
of disk star formation and chemical evolution are based on analytical
calculations  \cite{efstathiou00,silk01,ferreras01,silk03,matteucci06,naab06}.
 This paper extends the analytic approach to study  
star formation histories that vary
systematically along the Hubble sequence. We  study  global
star formation, incorporating supernova feedback, gas accretion
and enriched outflows in disks modelled by a  multi-phase interstellar medium 
in a fixed gravitational potential. 

One of the current problems afflicting galaxy formation models is the
role of gas infall. For both ellipticals \cite{bower06,croton06} and massive 
disks at $z\sim 2$ \cite{forster06},
infall rates are sufficiently high from CDM theory (ellipticals and disks) and
observations (disks) that infall must be quenched, otherwise distant  
ellipticals are too blue and the disks are too massive by the current 
epoch. The problems may be related via feedback from AGN, but the details 
are poorly understood with regard to the resulting star formation history, 
gas infall rates and chemical evolution. In this paper, we focus on disk 
galaxies, and develop a phenomenological description of disk evolution in 
which the onset and duration of the gas infall history are found to
be the controlling parameters. The star formation histories, gas distributions 
and chemical evolution can be explained in a simple sequence of
models which are primarily regulated by the cold gas accretion history.

One of the main features of galaxies that characterises the significance 
of the  Hubble sequence is the 
wide range
in young stellar content and star formation activity. This variation in
stellar content is part of the basis of the Hubble classification itself, and
understanding its physical nature and origins is fundamental to understanding
galaxy evolution in its broader context. In this paper, we construct 
a sequence of evolutionary models in which 
present-day  properties of different types of disk galaxies are reproduced.

The first implementation of supernova-driven feedback in the context of CDM was
to account for the properties of dwarf galaxies \cite{dekel86}, and
subsequent studies have explored the role of supernova feedback in more
massive galaxies.
Feedback is an important element
in our attempts to model galaxy  evolution. Energy injection from supernovae
is probably the most plausible feedback mechanism for systems with virial
temperatures higher than $10^{5}$ K. Winds from quasars might also disrupt
galaxy formation or limit the growth of central black holes. Here, we will be
concerned exclusively with supernova-driven feedback and we will not
consider feedback from an active nucleus. The model presented in this paper
is similar to the simple self-regulating model with inflow and outflow
developed in \pcite{efstathiou00}, the main difference being  in the infall
model 
that we implement.We adopt an exponentially
 decreasing infall rate normalised in
 order to reproduce the observed total disk mass density in the solar
neighbourhood, whereas in Efstathiou's model it is the conservation of
specific angular momentum that specifies the final radius in the disc for
each gas element. Angular momentum conservation is known to be a poor
approximation, at least in the numerical simulations, and our more
phenomenological model is  easily adapted to the chemical evolution 
constraints.

The main result of this paper is that star formation histories of the
different types of disk galaxies can be reproduced using a one parameter
model. The key parameter of our model is the time corresponding to the onset
of the  infall. Using this free parameter, we reproduce the distribution of disk
birthrate parameters $b$, the ratio of the current SFR to the average past SFR,
for each type of disk galaxy as presented in \pcite{kennicutt94} and summarised in Figure~\ref{fig:FitGauussian}.
We then explore the various implications of the model for the radial and
temporal dependences of the gas fraction, star formation rate and
metallicity.

The layout of this paper is as follows. Section~\ref{sec:Model} briefly
reviews the main points of our model and presents some basic results.
The model is extended in Section~\ref{sec:Refinements} to include an
improved treatment of  winds and chemical evolution. Finally, in
Section~\ref{sec:discuss} we discuss our results and present our conclusions.

\section{Model}
\label{sec:Model}
The model described here is a self-regulating model with inflow and
outflow. The disk is considered to be a system of independent rings each of
$35{\times}{r_{d}}$~pc wide (where $r_{d}$ is in units of kpc)\footnote{For a disk galaxy with scale length $r_{d}=~3$kpc the width of each zone is equal to $35\times3=~105$ pc}. Neither radial inflows nor radial outflows are
considered. The ring centred at the galactocentric distance
$r_{\odot}=~8.5$~kpc is labelled as the solar neighbourhood. For the
present-day 
total disk surface density in the solar neighbourhood, we adopt a value
${\Sigma}_{tot}(r_{\odot},t_{g})=~60$ M$_{\odot}$pc$^{-2}$ (\pcite{holmberg04}
found $56{\pm}6$~M$_{\odot}$pc$^{-2}$ for their disk model).

\subsection{Dark Halo and Disk Model}

The dark halo is assumed to be described by the \pcite{navarro95} profile,

\begin{equation}
\rho(r)=~{\frac{{\delta}{\rho}_{c}}{(Cx)(1+Cx)^{2}}}~,
\end{equation}
where $x{\equiv}r/r_{v}$, ${\rho}_{c}$ is the critical density, $r_v$ is the
virial radius at which the halo has mean overdensity of 200 with respect to
the background and $C$ is the concentration parameter. For the present model, we adopt
a value for $C$  of 10. The circular speed corresponding to this profile is
\begin{equation}
{v_{H}}^{2}(r)=~v_{v}^{2}{\frac{1}{x}}{\frac{[ln(1+Cx)-Cx/(1+Cx)]}{[ln(1+C)-C/(1+C)]}}~,
\end{equation}
where $v_{v}^{2}{\equiv}{\frac{GM_{v}}{r_{v}}}$,~and~$M_{v}$ is the mass of
the halo within the virial radius.

For the surface density of the disk at $t=~0$, we assume an exponential profile:
\begin{equation}
{\Sigma}_{tot}(r,0)=~{\Sigma}_{0}e^{\frac{-r}{r_{d}}},
\end{equation}
where $M_{D}=~2{\pi}r_{d}^{2}{\Sigma}_{0}$, ${\Sigma}_{tot}(r,0)$ is the total
surface density of the ring centred at galactocentric radius $r$ at t=~0, $M_{D}$ is
the total disk mass at t=~0 and $r_{d}$ is the scale length. So in our model the entire gas disk has formed instantaneously at $t=~0$. The rotation curve of a
cold exponential disk is given by \cite{freeman70}:
\begin{equation}
{v_{D}}^{2}(r)=~2v_{c}^{2}y^{2}[I_{0}(y){K}_{0}(y)-I_{1}(y)K_{1}(y)]
\end{equation}
where $y{\equiv}{\frac{1}{2}}{\frac{r}{r_{d}}}$ and
$v_{c}^{2}{\equiv}\frac{GM_{D}}{r_{d}}$. For ratio $\frac{v_{v}}{v_{c}}$ we adopt a value equal to 0.45 that corresponds to the Milky Way. Models using standard disk formation theory with adiabatic contraction within the cuspy halo~\cite{klypin02} reproduce the broad range of observational data available for the Milky Way by adopting a virial mass equal to M$_{v}~{\approx}~10^{12}$M$_{\odot}$, baryonic mass equal to M$_{bar}~{\approx}~4-6{\times}10^{10}$M$_{\odot}$ and virial radius r$_{v}=~258$~kpc. These values in combination with an adopted value for the disk scale length equal to r$_{d}=~3$~kpc~\cite{sackett97} give $\frac{v_{v}}{v_{c}}=~0.45$. In this work, we keep the ratio $\frac{v_{v}}{v_{c}}$ constant for all the models we present.\footnote{The ratio $\frac{v_{v}}{v_{c}}$ defines the disk scale length $r_{d}$. The disk scale length is $r_{d}=~\frac{(\frac{3M_{D}}{4{\pi}200{\rho}_{c}f_{coll}(\frac{v_{v}}{v_{c}})^{2}})^{1/3}}{f_{coll}}$~kpc}. The ratio of virial radius of the halo to disk scale length  is defined as the collapse factor
$f_{coll}{\equiv}{\frac{r_{v}}{r_{d}}}=~50$; such a value is needed to reproduce a median value of $\approx 0.05$ for the dimensionless spin parameter of the halo $\lambda_{H}$  \cite{efstathiou00}. The disk is
truncated at radius $r/r_{d}\approx7.$

\subsubsection{Two-component(stellar and gas) rotating disk}

The stellar radial velocity dispersion ${\sigma}_{\star}$ is related to that
of the gas clouds:
\begin{equation}
{\sigma}_{\star}=~{\alpha}{\sigma}_{g},
\end{equation}
Here we assume ${\alpha}=~5$ \cite{efstathiou00}. The scaleheight for a 
two-component rotating disk is~\cite{talbot75}:
\begin{equation}
H=~\frac{{{\sigma}_{g}}^{2}}{{\pi}G{\Sigma}_{g}}\frac{1}{(1+\frac{{\Sigma}_{\star}/{\Sigma}_{g}}{{\sigma}_{\star}/{\sigma}_{g}})}
\end{equation}

\subsection{Epicyclic frequency Model}
The epicyclic frequency is given by the expression
\begin{equation}
{\kappa}=~2\omega(1+{\frac{1}{2}}{\frac{r}{\omega}}{\frac{d{\omega}}{dr}})^{1/2},\end{equation}
in which $\omega$ is the angular velocity of the disk. In the above equation
we replace $\omega=~v_{tot}/r$ where $v_{tot}^{2}=~v_{H}^{2}+v_{D}^{2}$. So the epicyclic
frequency(in units of $10^{-15}$sec$^{-1}$) is:
\begin{equation}
{\kappa}=~0.035\sqrt{2}{\frac{v_{tot}}{r}}(1+{\frac{r}{v_{tot}}}{\frac{dv_{tot}}{dr}})^{1/2}
 \end{equation}

\subsection{Star Formation Rate(SFR)}

In this work we adopt the star formation law proposed in \pcite{wang94}, to
which we refer for a detailed description. The star formation rate (in units
of M$_{\odot}$pc$^{-2}$Gyr$^{-1}$) is given by :
\begin{equation}
{\psi}(r,t)=~\frac{{\epsilon}{\kappa}{\Sigma}_{g}(1-Q^{2})^{1/2}}{Q}
\end{equation}
where ${\kappa}$ is the epicyclic frequency, ${\Sigma}_{g}$ is the gas
surface density in units of M$_{\odot}$pc$^{-2}$, $Q$ is the gravitational
instability parameter\cite{toomre64} and $\epsilon$ is the efficiency of star
formation. For the model in this paper we set $\epsilon=~0.02$.

The star formation rate defined above has the following features: (i) star
formation can only occur if the disk is gravitationally unstable,
$Q<1$; (ii) the degree of instability of the disk, measured by $Q$, is directly
linked to the star-formation rate; the smaller $Q$ is, the more rapidly stars
are formed.

The stability criterion for a two-component (stellar and gas) rotating disk
can be written as~\cite{wang94}:
\begin{equation}
Q=~\frac{\kappa}{{\pi}G}(\frac{{\Sigma}_{g}}{{\sigma}_{g}}+\frac{{\Sigma}_{\star}}{{\sigma}_{\star}})^{-1}
\end{equation}
where ${\Sigma}_{\star}$ and ${\sigma}_{\star}$ are the stellar surface
density (in units of M$_{\odot}$pc$^{-2}$) and the radial velocity dispersion
(in units of $\rm  km\,sec^{-1}$), and~${\kappa}$ is the epicyclic frequency.

\subsection{IMF}

The adopted stellar initial mass function is of the standard Salpeter form :
\begin{equation}
\frac{dN{\star}}{dm}=~Bm^{-(1+x)},~m_{l}<m<m_{u},~x=~1.35
\end{equation}
where $m_{l}=~0.1$ M${_{\odot}}$,$m_{u}=~50$M${_{\odot}}$.

For the IMF adopted, one supernova is formed for every $125M{_{\odot}}$ of
star formation, assuming that each star of mass greater than 8$M_{\odot}$
releases $10^{51}E_{51}$ ergs in kinetic energy in a supernova
explosion. Therefore, the energy injection rate per unit surface area (in units
of erg~s$^{-1}$pc$^{-2}$) is given by:

\begin{equation}
{\dot{E}}_{sn}^{\Omega}=~2.5{\times}10^{32}E_{51}{\epsilon_{c}}{\psi}
\end{equation}
where the parameter $\epsilon_{c}$ defines the percentage of supernovae
energy that goes to the ambient medium. For this model we adopt $\epsilon_{c}=~0.03$.

\subsection{Energy dissipation due to cloud inelastic collisions}
The rate of energy loss per unit surface area (in units of
ergs$^{-1}$pc$^{-2}$) is given by~\cite{efstathiou00}:
\begin{equation}
{\dot{E}}_{coll}^{\Omega}=~5.{\times}10^{29}(1+\frac{{\Sigma}_{\star}/{\Sigma}_{g}}{{\sigma}_{\star}/{\sigma}_{g}}){\sigma}_{5g}{\Sigma}_{5g}^{3}.
\end{equation}
Using the above equations we know the energy balance in each ring of the
disk. We define the radial velocity dispersion of clouds (in units of
$\rm km\,s^{-1}$) as:
\begin{equation}
{\sigma}_{g}(r,t)=~\sqrt{\frac{2{\times}E(r,t)}{M_{g}(r,t)}}
\end{equation}
where ${E}(r,t)$ is the energy balance at galactocentric radius $r$ and time $t$,
and~$M_{g}(r,t)$ is the mass of the cold gas at galactocentric radius $r$ and
time $t$.

In summary, the energy balance in each ring of the disk defines the cloud and
stellar radial velocity dispersion ${\sigma_{g}}, {\sigma}_{\star}$ through
equation (14) and equation (5). Combining this piece of information with the
cloud and stellar surface densities ${\Sigma_{g}}, {\Sigma}_{\star},$ we
calculate the degree of instability at galactocentric radius $r$ through
equation(10). The star formation rate is regulated by the degree of
instability, so the final form of the star formation rate is derived by
inserting equation (10) into equation (9).

\subsection{Infall Model}
\label{sec:Infall Model}

The adopted form for the gas infall rate is an exponentially decreasing
function in which the rate of gas infall (in units of
M$_{\odot}$pc$^{-2}$Gyr$^{-1}$) in each ring is expressed as:

\begin{equation}
f(r,t)=~A(r)e^{-{\frac{t}{{\tau}_{f}}}}
\end{equation}
where ${\tau}_{f}$ (in units of Gyr)~is the infall time scale. The infall rate
$f(r,t)$ is normalised to the present-day local disk density,
${{\int}_{t_{low}}^{t_{g}}}f(r,t)dt=~{\Sigma}_{tot}(r,t_{g})$, where
${\Sigma}_{tot}(r,t_{g})$ is the present-day total disk surface density of
the ring centred at galactocentric radius $r$,~and~$t_{g}$ is the age of the
galactic disk ($t_{g}=~10$ Gyr). Assuming that the present-day total masses of
the different rings exponentially decrease with increasing galactocentric
distance with scale-length $r_{d}$, the form of $A(r)$ can be written as
\cite{chang02}:

\begin{equation}
A(r)=~\frac{{\Sigma}_{tot}(r{_{\odot}},t_{g})-{\Sigma}_{tot}(r{_{\odot}},0)}{{{\int}_{t_{low}}^{t_{g}}}e^{-{\frac{t}
{{\tau}_{f}}}} dt}e^{\frac{-(r-r{_{\odot}})}{r_{d}}}
\end{equation}
where ${\Sigma}_{tot}(r{_{\odot}},0)$ is the total disk surface density in
the solar neighbourhood at the epoch which the formation of the disk begins,
and ${\Sigma}_{tot}(r{_{\odot}},t_{g})$ is the present-day total disk surface
density in the solar neighbourhood. We adopt an exponentially decreasing infall rate normalised in order to reproduce the observed total mass density in the solar neighbourhood. In our infall model, we have introduced a
parameter $t_{low}$ that corresponds to the time that infall switches
on. Adjusting this parameter we can reproduce the distribution of disk
birthrate parameter {\it b}, the ratio of the current SFR to the average past SFR,
for each type of galaxy presented in \pcite{kennicutt94}. The infall time
scale ${\tau}_{f}$ (in units of Gyr) is also a free parameter of the model but
we keep it constant, adopting a value of 0.5 Gyr.

According to the adopted infall form, since ${\tau}_{f}$ does not vary with radius, the disk keeps the exponential form with scale length $r_{d}$ at each epoch of evolutionary time.
\subsection{Hot Phase Model}

An expanding supernova remnant will evaporate a mass of cold gas. The rate of
the evaporated mass per unit area is (Efstathiou 2000):

\begin{equation}
{\dot{M}}_{ev}^{\Omega}{\approx}1{\times}10^{-7}{\frac{{\sigma}_{5g}^{2}}{{\Sigma}_{5g}(1+\frac{{\Sigma}_{\star}/{\Sigma}_{g}}{{\sigma}_{\star}/{\sigma}_{g}})}}{\times}S_{13}^{0.71}{\gamma}^{0.29}E_{51}^{0.71}f_{\Sigma}^{-0.29}
\end{equation}
in units of M$_{\odot}$pc$^{-2}$yr$^{-1}$, where ${\Sigma}_{5g}$ is the gas
surface density in units of 5~M$_{\odot}$pc$^{-2}$, ${\sigma}_{5g}$ is the
cloud radial velocity dispersion in units of 5~kms$^{-1}$ and $S_{13}$ is the
 supernovae rate in units of $10^{-13}$pc$^{-3}$yr$^{-1}$. The parameter ${\gamma}$ relates the blast wave velocity to the isothermal sound 
speed (v$_{b}=~{\gamma}c_{h}$, ${\gamma}~{\approx}~2.5$). The parameter $f_{\Sigma}$ 
is defined through the evaporation parameter~(in units of pc$^{2}$):
\begin{equation}
{\Sigma}^{ev}=~280\frac{ {{\sigma}_{5g}}^{2} } {{{\Sigma}_{5g}}^{2}
}\frac{1}{(1+\frac{ \frac{{\Sigma}_{\star}}{{\Sigma}_{g}}}
{\frac{{\sigma}_{\star}}{{\sigma}_{g}}})}\frac{1}{{\phi}_{\kappa}}=~f_{\Sigma}{{\Sigma}^{ev}}_{\odot}
\end{equation}
where ${{\Sigma}^{ev}}_{\odot}$ is the evaporation parameter in the local solar
neighbourhood and ${\phi}_{\kappa}$ is a parameter that quantifies the
effectiveness of classical thermal conductivity. For this model we adopt
${{\Sigma}^{ev}}_{\odot}=~95$ pc$^{2}$ and ${\phi}_{\kappa}=~0.1$. For a system with
porosity Q close to unity, the temperature and the age of a supernovae
remnant are given by:

\begin{eqnarray} 
t_{o}=~5.5{\times}10^{6}{S_{13}}^{-5/11}{\gamma}^{-6/11}E_{51}^{-3/11}n_{h}^{3/11}
{\rm yr}, \\
T_{o}=~1.2{\times}10^{4}S_{13}^{6/11}{\gamma}^{-6/11}E_{51}^{8/11}n_{h}^{-8/11}
{\rm K},
\end{eqnarray}
The density of the ambient hot phase is
\begin{equation}
n_{h}=~4.3{\times}10^{-3}S_{13}^{0.36}{\gamma}^{-0.36}E_{51}^{0.61}f_{\Sigma}^{-0.393}
{\rm cm^{-3}}.
\end{equation}
For $10^{5}{\leq}T{\leq}10^{6} K$ we adopt a cooling rate of ${\Lambda}
\approx 2.5{\times}10^{-22}$T$_{5}^{-1.4}$ ergcm$^{3}$s$^{-1}$. So for a gas with
primordial composition, the cooling time $t_{cool}$ is
\begin{equation}
t_{cool}=~2t_{o}T_{5}^{2.4}f_{\Sigma}^{0.5}
\end{equation}

\subsection{Outflow Model}
\label{sec:Outflow Model}

As mentioned in the previous section, supernova bubbles expand, evaporate
cold gas and compress the ambient interstellar medium(ISM). The compressed ISM
will also be driven by a form of wind. In order to study the effects of
outflow, we assume a simple phenomenological model for galactic winds where
the wind mass-loss rate ${\dot{M}}_{W}$ (in units of
M$_{\odot}$pc$^{-2}$Gyr$^{-1}$) is assumed to be proportional to the star
formation rate.
\begin{equation}
{\dot{M}}_{W}=~k{\psi}~.
\end{equation}
We explore many values of k between 0 and 1. The hot gas that escapes from the halo is removed permanently. Superwinds indeed are found around Milky Way-type galaxies
\cite{str07}.  This is puzzling from the theoretical perspective
because supernovae are considered to be incapable of driving a strong
wind from the gravitational potential well of a massive galaxy.
 Also, at least one example is known of an extended
x-ray halo around a normal massive spiral galaxy \cite{ped06}. In this
case, gravitational accretion and shock heating provides the most
plausible source for heating the gas.

\subsection{Galactic Fountain Model}
Models of gas flow on the galactic scale were first introduced by \pcite{shapiro76} and subsequently developed by \pcite{bregman80} and others. The galactic fountain originates from the supernovae that warms up the disk gas to temperatures of $10^{6}$K. The upflowing gas cools and condenses into neutral hydrogen clouds that rains onto the disk. The models assume the height to which the hot gas will rise and the expected rate of condensation in the cooling gas depends only on the temperature of the gas at the base of the fountain and the rate of cooling of the upflowing gas.
In the model described in this paper the hot gas that is not lost from the disk returns to the disk at the radius from which it was expelled after a time $t_{cool}$.

\subsection{Chemical Evolution}
\label{sec:Chemical Evolution}
We include chemical evolution in the model using the instantaneous recycling
approximation. So we assume that all processes involving stellar evolution,
nucleosynthesis and recycling take place instantaneously on the time scale of
galactic evolution. The equation of galactic chemical evolution is
\begin{equation}
{\Sigma}_{g}dZ=~p{\psi}dt+(Z_{F}-Z)f
\end{equation}
where f is the infall rate (in units of M$_{\odot}$pc$^{-2}$Gyr$^{-1}$) and p
is the yield \cite{pagel97}. We adopt a yield of $p=~0.02$ and assume that the
mass accreted to the disk has zero metallicity ($Z_{F}=~0$).We normalise the
metallicities to the solar value for which we adopt $Z_{\odot}=~0.02$.

\subsection{Different Types of Spiral Galaxies}
To explore the evolution of different types of spiral galaxies we adopt five
models in this paper (See Table~\ref{Table.1}). These models differ only in
the mass of the disk and the disk scale length $r_{d}$. All the input parameters that these models have in common are summarized in Table~\ref{Table.Input}.

\begin{table}
  \begin{center}

  \begin{tabular}{|l|c|}
    \hline 
    Input~parameters & \\ 
    \hline

    Solar radius&$r_{\odot}=~8.5$~kpc\\ 
    Total disk surface density at $r_{\odot}$&${\Sigma}_{tot}(r_{\odot},t_{g})=~60$ M$_{\odot}$pc$^{-2}$\\ 
    Star Formation IMF&Salpeter\\
    Solar metallicity&$Z_{\odot}=~0.02$\\ 
    Effective yield&$p=~0.02$\\
    Star formation efficiency&$\epsilon=~0.02$\\
    Infall timescale&$\tau_{f}=~0.5$~Gyr\\
    Galactic disk age&$t_{g}=~10$~Gyr\\
    Thermal Conductivity effectiveness&$\phi_{\kappa}=~0.1$\\
    Solar evaporation parameter&${{\Sigma}^{ev}}_{\odot}=~95$~pc$^{2}$\\ 
    Disk to Halo velocity ratio&$\frac{v_{v}}{v_{c}}=~0.45$\\
    Collapse factor&$f_{coll}=~50$\\
    Wind Model Parameter&$k=~0.2$\\
    \hline
  \end{tabular}
\end{center}
\caption{Input parameters in common for the five models we examine in this paper.}
\label{Table.Input}
\end{table}

\begin{table}
  \begin{center}

  \begin{tabular}{|c|c|c|}
    \hline 
    $Type$ & Disk Mass $M_{D}$($M_{\odot}$) & $r_{d}(kpc)$ \\ 
    \hline

    Sa&$5.5\times10^{10}$&3.\\ 
    Sb&$4.\times10^{10}$&2.7\\
    Sc&$2.7\times10^{10}$&2.37\\ 
    Sd&$1.5\times10^{10}$&1.95\\
    Sm&$1.\times10^{10}$&1.7\\ 
    \hline
  \end{tabular}
\end{center}
\caption{Five models adopted in this paper}
\label{Table.1}
\end{table}

\subsection{Basic Results}
In this section will present the basic results of the model discussed in the
previous sections. Unless otherwise stated, we adopt a value 0.2 for the
parameter k in the outflow model. For clarification of the plots, when results from models Sc and Sd are similar we define a new type of disk galaxy Sc/d. The  data for Sc/d type come from the average between the Sc model data set and the Sd model data set.
\subsubsection{Distribution of disk birth rate b}
Values of the birth parameter {\it b} were calculated for each type of spiral
galaxy. For the calculation of {\it b} values we ignore the SFR that corresponds to 
the first $10^{8}$yr, due to a deficiency in our model. More specifically in
the model we assume that the entire gas disk has formed instantaneously at
$t=~0$: this is unrealistic and leads to very high rates of star formation for
the first $10^{8}$yr. Figure~\ref{fig:birthrate} shows the disk birthrate
parameter {\it b} subdivided by galaxy type as a function of our free
parameter $t_{low}$ that corresponds to the time that infall switches on. The
dotted line indicates the median value of birthrate parameter while the dashed lines show the 
quartiles of birthrate parameter presented in \pcite{kennicutt94}. In Table~\ref{Table.2}, we 
present the value of the free parameter $t_{low}$ we adopt in order to reproduce the median value of
birthrate parameter for each type of disk galaxy.

\begin{table}
  \begin{center}

  \begin{tabular}{|c|c|c|}
    \hline $Type$ & b(median) & $t_{low_{med}}(Gyr)$ \\ \hline

    Sa& $ 0.07$ &0.10\\ Sb& $ 0.37$ &7.00\\ Sc& $ 1.04$ &8.05\\ Sd& $ 0.63$
    &7.80\\ Sm& $ 1.31$ &8.35\\ \hline
  \end{tabular}
\end{center}
\caption{Median value of {\it b} and $t_{low}$ value for different models.}
\label{Table.2}
\end{table}

\begin{figure}	
\epsfig{file=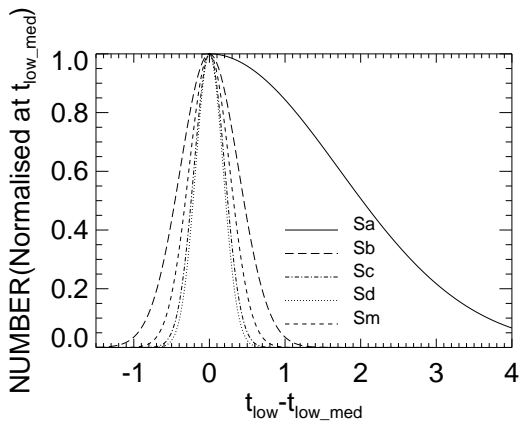,width=8cm}
\caption{Gaussian fits to the {\it b} parameter dispersion for each type of disk
galaxy}
\label{fig:FitGauussian}
\end{figure}

\begin{figure*}	
\epsfig{file=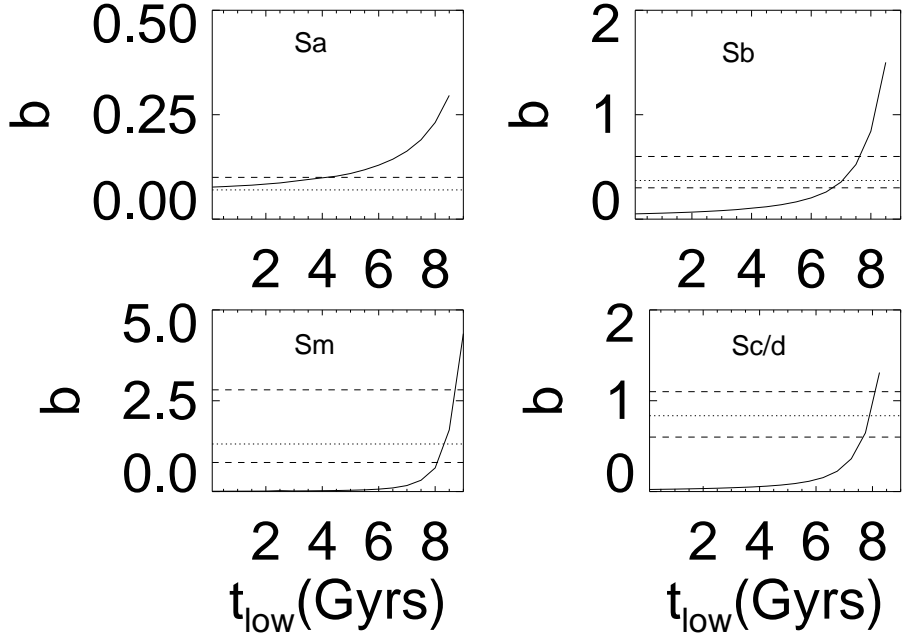,width=12cm}
\caption{The disk birthrate parameter {\it b}, the ratio of current SFR to
the average past SFR, plotted against the free parameter $t_{low}$, for
different types of disk galaxies. The dotted line indicates the median values
of birthrate parameter while the dashed lines show the quartiles of birthrate parameter presented 
by Kennicutt et al. For clarification of the plot, we define a new type of disk 
galaxy Sc/d. The  data for Sc/d type come from the average between the Sc model data set and the Sd 
model data set.}
\label{fig:birthrate}
\end{figure*}

To make our results more useful for observers, we fit Gaussians to the {\it
b} parameter dispersion for each type of disk galaxy
(Figure~\ref{fig:FitGauussian}). The full width at half maximum for each
different Gaussian equals the width of the $t_{low}$ parameter distribution
in order to reproduce the quartiles of the {\it b} parameter distribution
presented in \pcite{kennicutt94}. In Table~\ref{Table.3}, we give the value
of ${\sigma}$ for the Gaussian fits as well as the width in $t_{low}$ in
order to reproduce the quartiles of the {\it b} parameter distribution. The
width of the {\it b} parameter distrbution between the quartiles for each
type of disk galaxy is presented in the last column.

\begin{table}
  \begin{center}

  \begin{tabular}{|c|c|c|c|}
    \hline $Type$ & ${\sigma}$ & ${\Delta}t_{low}(Gyr)$ & ${\Delta}b$\\
    \hline

    Sa& $ 1.716$ &4.04 &0.1\\ Sb& $ 0.386$ &0.91 &0.3\\ Sc& $ 0.205$ &0.48
    &0.6\\ Sd& $ 0.183$ &0.43 &0.4\\ Sm& $ 0.270$ &0.64 &2.0\\ \hline
  \end{tabular}
\end{center}
\caption{The value of ${\sigma}$ for the Gaussian fits (second column), the
width in $t_{low}$ in order to reproduce the quartiles of the {\it b}
parameter distribution (third column), the width of the {\it b} parameter
distribution between the quartiles of the distributions (fourth column). The
values are shown for different types of disk galaxies.}
\label{Table.3}
\end{table}

\subsubsection{Density Profiles}

\begin{figure*}	
\epsfig{file=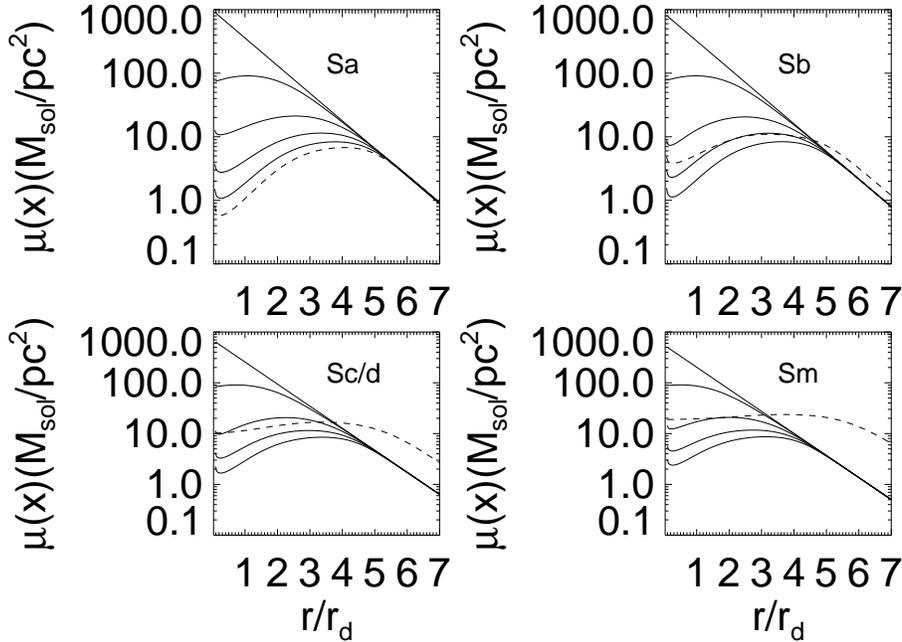,width=12cm}
\caption{The evolution of the gas surface densities for different ages and
different types of disk galaxies. The results are shown for ages (top - bottom) of 0, 0.1, 1,
3, 6 and 10 Gyr ({\it dashed} line).}
\label{fig:Gas_Density_Prof}
\end{figure*}

\begin{figure*}	
\epsfig{file=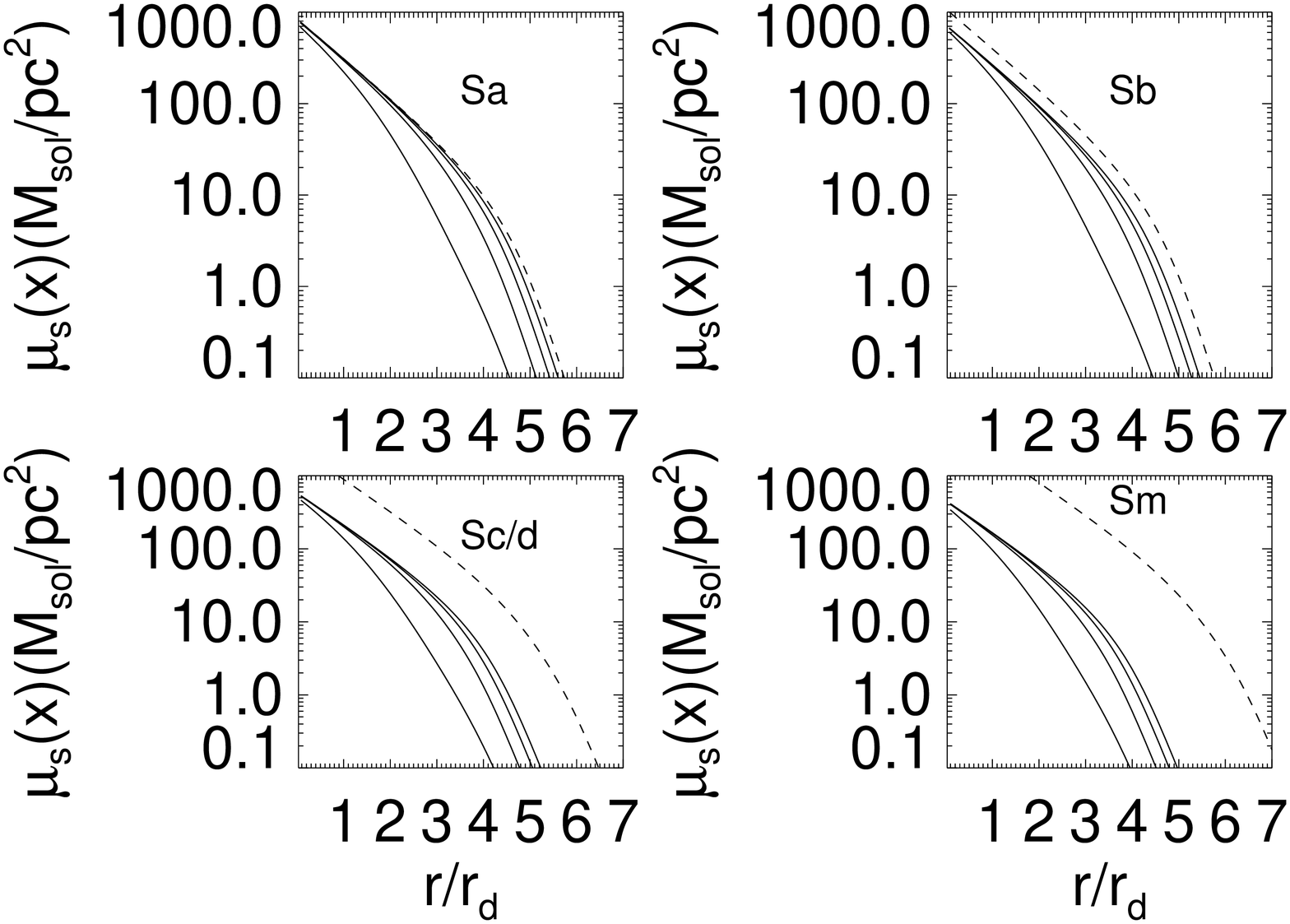,width=12cm}
\caption{The evolution of stellar surface densities for different ages and
different types of disk galaxies. The results are shown for ages (top - bottom) of 0, 0.1, 1,
3, 6 and 10 Gyr ({\it dashed} line).}
\label{fig:Stars_Density_Prof}
\end{figure*}

Figure~\ref{fig:Gas_Density_Prof} and figure~\ref{fig:Stars_Density_Prof}
show the evolution of the gas and stellar surface densities, respectively, (in
units of M$_{\odot}$pc$^{-2}$) for different ages and different type of disk
galaxies. The results are shown for ages of 0, 0.1, 1, 3, 6 (top - bottom) and 10 Gyr ({\it dashed} line). The star formation rates are initially high and hence 
the time scale for star formation is short. Almost the total amount of gas in 
the central region of disk is transformed to stars in $10^{8}$yr. The star formation rate declines rapidly after 1~Gyr. As figure~\ref{fig:Gas_Density_Prof} and figure~\ref{fig:Stars_Density_Prof} show the star formation at early times is concentrated to the inner parts of the disk which have a high surface density, and hence the gas distribution develops a surface density profile with an inner 'hole', similar to what is seen in the HI distributions in real galaxies \cite{deul90}. 

Model Sa has parameters similar to those of the Milky Way. Gas surface densities from direct observations at the solar neighbourhood is $8{\pm}5$ M$_{\odot}$/pc$^{2}$~\cite{dame93} in good agreement with the predictions of the model. Slightly higher values of ${\approx}~13-14$ M$_{\odot}$/pc$^{2}$ have been reported by \pcite{olling01} and are in good agreement with Sb model. Taking into account the scatter in the measurements of Milky Way scalelength~($r_{d}\approx2.5-3.5$kpc \pcite{sackett97} ) Sb model can also be a good candidate for the Milky Way. 

Direct observations of the stars at the solar neighbourhood of Milky Way by \pcite{gould06} and \pcite{zheng01} using HST observations of M stars in the Galactic disk determine a surface density of $12.2-14.3$ M$_{\odot}$/pc$^{2}$. Visible stars other than M stars contribute  ${\approx}~15$ M$_{\odot}$/pc$^{2}$, resulting in a total stellar surface density in the range of ${\approx}~27-30$ M$_{\odot}$/pc$^{2}$. This value is in good agreement with predictions from Sa and Sb models.

\subsubsection{Gas Velocity Profiles}
\begin{figure*}	
\epsfig{file=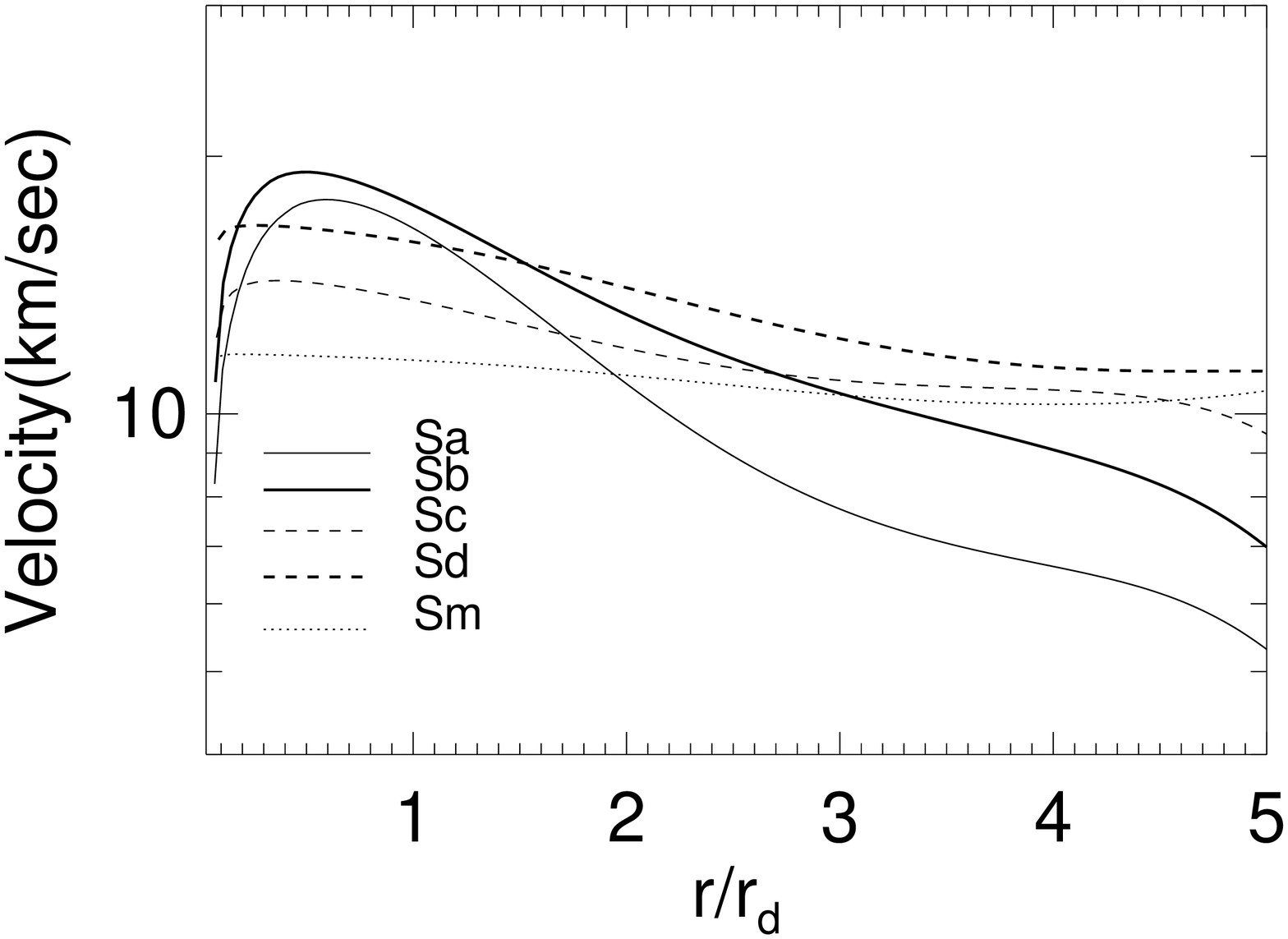,width=12cm}
\caption{The radial distribution of the gas velocity at 10 Gyr for different
types of disk galaxies, Sa~({\it solid} line), Sb~({\it thick solid} line),
Sc~({\it dashed} line), Sd~({\it thick dashed} line) and Sm~({\it dotted}
line).}
\label{fig:Velocity_Prof}
\end{figure*}

Figure~\ref{fig:Velocity_Prof} shows the radial distribution of the gas
velocity at 10~Gyr for different types of disk galaxies, Sa~({\it solid} line),
Sb~({\it thick solid} line), Sc~({\it dashed} line), Sd~({\it thick dashed}
line) and Sm~({\it dotted} line). Gas velocity profiles are in good agreement
with recent theoretical predictions~\cite{ricotti02} that in a multiphase low metallicity~(Z${\approx}5\times10^{-3}Z_{\odot}$) interstellar medium  the velocity probability distribution should be close to a Maxwellian with velocity dispersion ${\sigma}{\ge}11$km/sec$^{-1}$. Furthermore studies of nearby face-on galaxies show that the velocity dispersion in the HI layer is decreasing monotonically from about $10-13$ kms$^{-1}$ in the optically bright inner regions to $6-8$ kms$^{-1}$ in the very outer parts \cite{kamphuis93}.

 \subsubsection{Star formation Profiles}
\begin{figure*}	
\epsfig{file=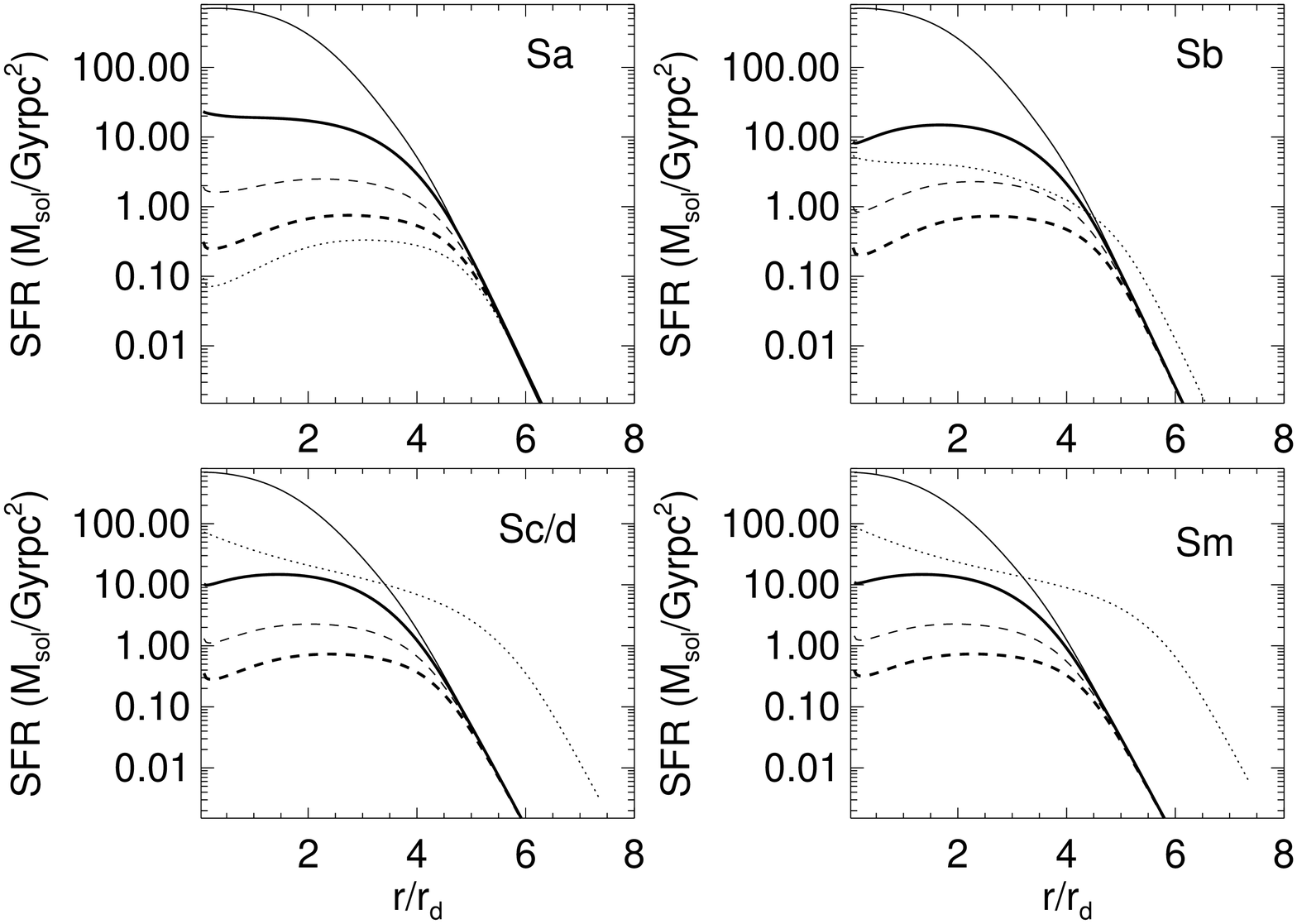,width=12cm}
\caption{The evolution of the radial distribution of the star formation rate
for different ages and different types of disk galaxies. The results are shown
for ages of 0.1 ({\it solid} line), 1 ({\it solid thick} line), 3 ({\it
dashed} line), 6 ({\it dashed thick} line) and 10 Gyr ({\it dotted} line).}
\label{fig:SFR_Prof}
\end{figure*}

Figure~\ref{fig:SFR_Prof} shows the evolution of the radial distribution of
the star formation rate for different ages and different types of disk
galaxies. The results are shown for ages of 0.1 ({\it solid} line), 1 ({\it
solid thick} line), 3 ({\it dashed} line), 6 ({\it dashed thick} line) and
10 Gyr ({\it dotted} line).

\subsubsection{Metallicity Profiles}
\begin{figure*}	
\epsfig{file=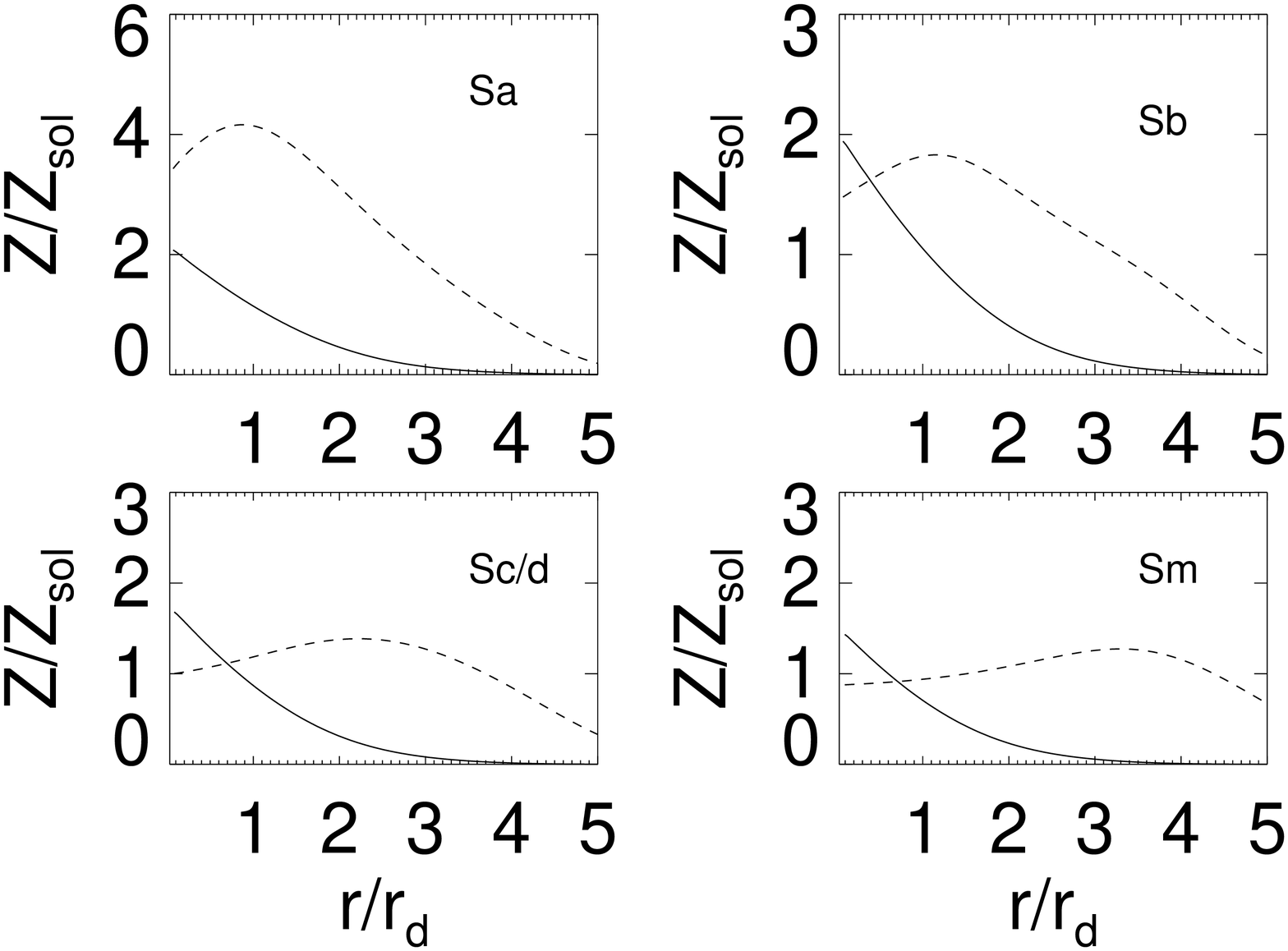,width=12cm}
\caption{The evolution of the radial distribution of gas metallicity for
different ages and different types of disk galaxies. The results are shown for
ages of 0.1 ({\it solid} line)and 10 Gyr ({\it dotted} line).}
\label{fig:Metallicity_Prof}
\end{figure*}

Figure~\ref{fig:Metallicity_Prof} shows the evolution of the radial
distribution of gas metallicity for different ages and different type of disk
galaxies. The results are shown for ages of 0.1 ({\it solid} line) and 10 Gyr
({\it dotted} line).

Note the very high value of metallicity for Sa type galaxies today. The model
predicts that metal-rich winds are needed especially for Sa type galaxies in
order to produce reasonable values for metallicity today. This prediction is
in agreement with suggestions presented in \pcite{dalcanton06} (for more
detailed comments see Section~\ref{sec:Improve Chemical
Evolution})\footnote{Note that metallicity today does decline slightly close
to the center. This is due to the radial profile we choose for the infall
rate since we add most of the fresh low metallicity gas to the central
regions.}.

\subsubsection{Infall Rate Profiles}
\begin{figure*}	
\epsfig{file=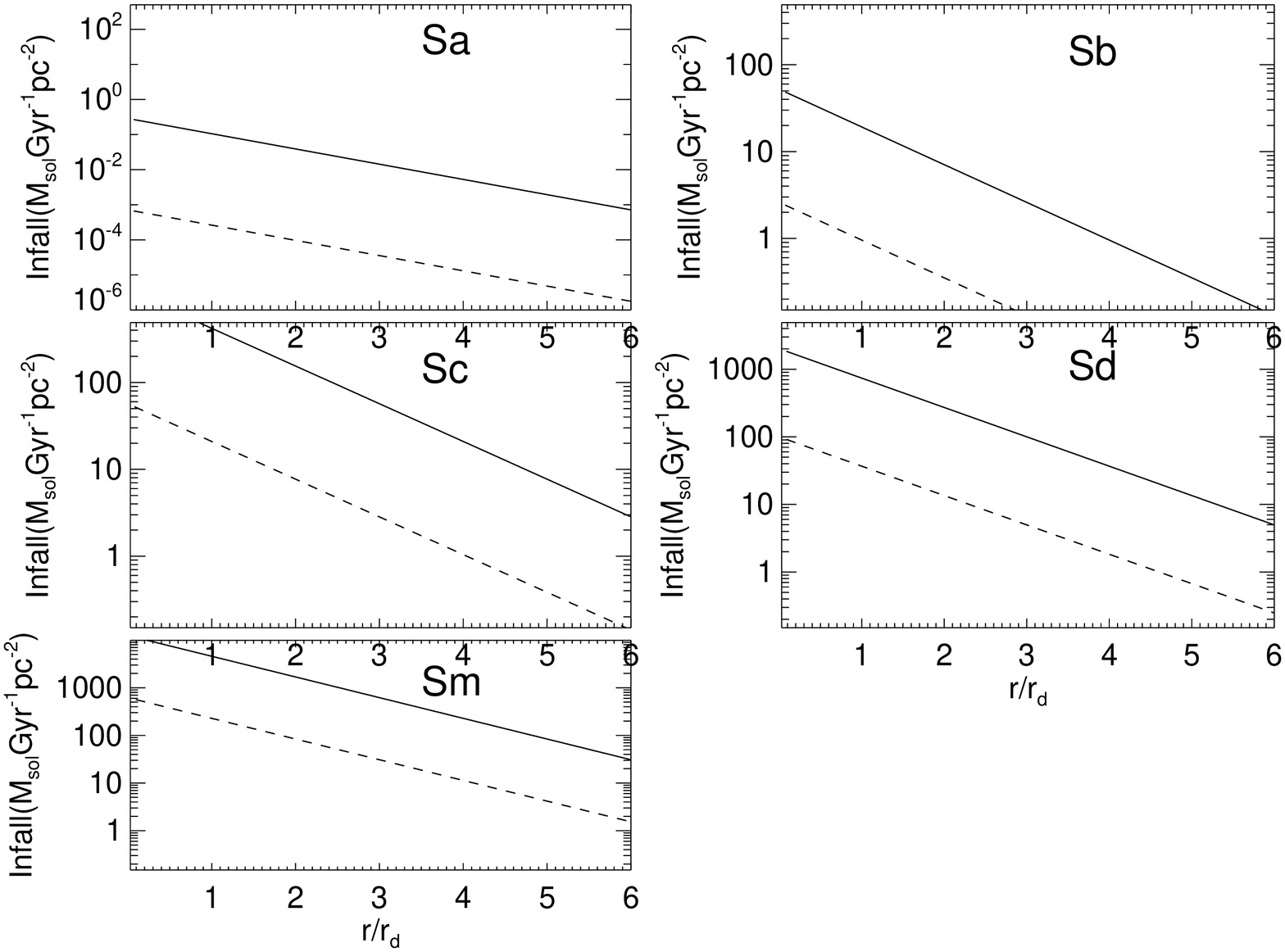,width=12cm}
\caption{The evolution of the radial distribution of the infall rate for
different ages and different types of disk galaxies. For Sa
galaxies the results are shown for 3 ({\it solid} line) and 6 Gyr ({\it
dashed} line). Infall rates for Sb, Sc, Sd, and Sm galaxies are presented for
ages of 8.5 ({\it solid} line) and 10 Gyr ({\it dashed} line)}
\label{fig:Infall_Prof}
\end{figure*}

The evolution of the radial distribution of the infall rate (in units of
M$_{\odot}$pc$^{-2}$Gyr$^{-1}$) is shown in Figure~\ref{fig:Infall_Prof}. The
results are shown for different ages. As already mentioned the time when the
infall switches on is defined by the free parameter $t_{low}$. For the
results presented in figure 5, we adopt for $t_{low}$ a value that reproduces
the median value of birth parameter {\it b} for each type of disk galaxy. For Sa
galaxies the results are shown for 3 ({\it solid} line) and 6 Gyr ({\it
dashed} line). Infall rates for Sb, Sc, Sd, and Sm galaxies are presented for
ages of 8.5 ({\it solid} line) and 10 Gyr ({\it dashed} line). Note the
different range in the y axis for different types of disk galaxies.

\subsubsection{Disk Global Properties}

\begin{figure*}	
\epsfig{file=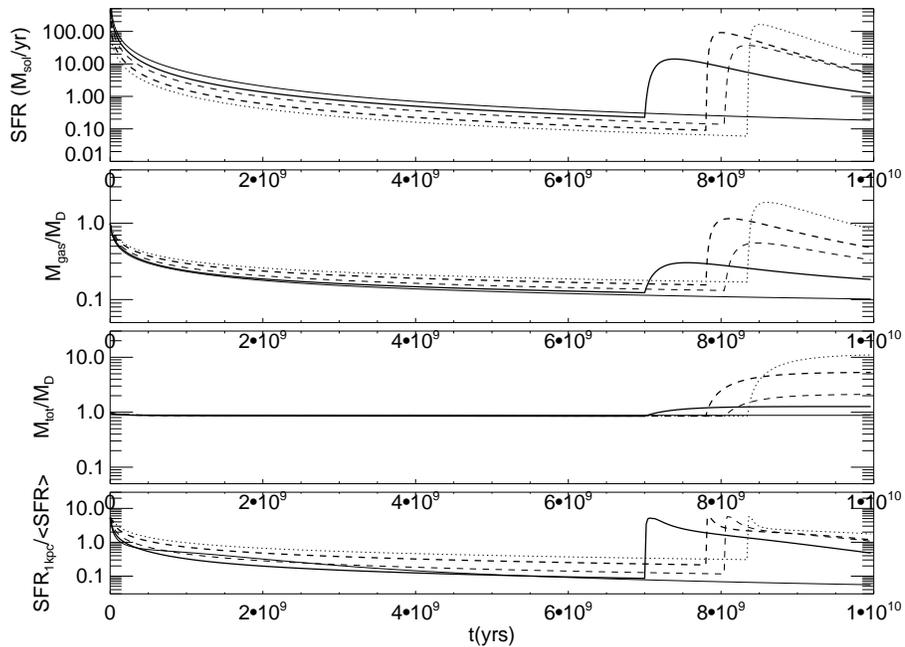,width=12cm}
\caption{Upper panel: time evolution of net star formation rate. Middle
panels: time evolution of gas fraction and total disk mass (stars plus gas). Bottom panel: ratio of star formation rate at $r=~1$~kpc over the mean star formation rate. The results are shown for different types of disk galaxies, Sa ({\it solid} line), Sb ({\it solid thick} line), Sc ({\it dashed} line), Sd ({\it dashed thick} line) and Sm ({\it dotted} line).}
\label{fig:Global_Prop}
\end{figure*}

Figure~\ref{fig:Global_Prop} shows the net star formation rates (upper panel)
and the evolution of gas fraction and total disk mass, stars plus gas (middle panels), for different types of disk galaxies respectively. Note the very high values for the infall rate for the Sb, Sc, Sd and Sm galaxies. Since these galaxies have smaller mass and the infall is switched on much later in comparison with the Sa type, we need high infall rates in order to reproduce the observed total mass density in the solar neighbourhood.

\subsection{Star formation rate histories}
In this section, we present results for the star formation rate history at
galactocentric radius $r=~1 \rm kpc$ for different types of disk galaxies.
\subsubsection{Galactocentric radius $r=~1$~kpc}
In figure~\ref{fig:Global_Prop}, we show the star formation rate history at
galactocentric radius$ $r equal to $r=~1$ kpc (bottom panel) for different types
of disk galaxies. Note, that galaxy types which are characterised by large
infall rates (Sb, Sc, Sd, Sm) show a peak at the onset of  infall.
 At time $t=~t_{low}$, when the infall switches on, the accreted gas
increases the local star formation rate by a bigger factor in comparison with the
increase in the mean star formation rate  across the disk and causes the peak in
Figure~\ref{fig:Global_Prop}. This phenomenon is much stronger at smaller
radii because of the inner 'hole' that has developed in the surface density
of the gas at late times (see figure~\ref{fig:Gas_Density_Prof}). At late
times, the star formation in the inner parts of the disk is stopped due to
lack of gas, but the fresh gas comes in at $t=~t_{low}$ to refuel the central
regions and increase the local star formation by a big factor. For Sa
galaxies, the infall rate is very small so the curve in
figure~\ref{fig:Global_Prop} is smooth.

\section{Refinements of the model}
\label{sec:Refinements}
The model described in the previous section contain a number of
simplifications, which we will attempt to refine in this section. We
introduce some simple improvements to the outflow
model~(Section~\ref{sec:Improve Outflow Model}) and to the chemical evolution
model~(Section~\ref{sec:Improve Chemical Evolution}).

\subsection{Outflow model}
\label{sec:Improve Outflow Model}

In Section~\ref{sec:Outflow Model} we assumed that the {\it k} parameter is
constant. Instead of keeping the {\it k} parameter in the outflow model
constant for the different types of disk galaxies, we can assume that {\it k}
is proportional to the ratio of bulge-to-disk circular velocities. This is a
reasonable assumption especially if we believe that AGN feedback contributes
to the wind mechanism. For systems with constant density, the bulge-to-disk
velocity ratio can be reduced to the following expression:
\begin{equation}
\frac{v_{b}}{v_{d}}{\propto}({\frac{r_{e}}{r_{d}}})^{1/a}
\label{eq:outflowprop}
\end{equation}
where $r_{e}$ is the effective radius of the bulge and $r_{d}$ is the the
disk scale length. Following \pcite{macarthur03}, we can use the following
expression for the ${\frac{r_{e}}{r_{disk}}}$ ratio :
\begin{equation}
{\frac{r_{e}}{r_{d}}}= 0.20 - 0.013(T - 5)~,
\label{eq:outflowmodel}
\end{equation}
where $T$ defines the galaxy type. The relationship between $T$ and galaxy type
is shown in Table~\ref{Table.T}. Note that eq.~\ref{eq:outflowmodel} is valid
only for galaxies with numerical type $T$ between $1$ and $7$. Initial
investigation of equation~\ref{eq:outflowmodel}, using $a=~1$ and $a=~2$, shows
that our results are insensitive to this refinement of the model. We plan to
investigate the implications of equation~(\ref{eq:outflowmodel}) in more
detail in a future paper.

\begin{table}
  \begin{center}

  \begin{tabular}{|c|c|c|c|c|c|}
    \hline de Vaucouleurs &Sa &Sb &Sc & Sd & Sm \\ \hline

    T&$1$&$3$&$5$&$7$&$9$\\

    \hline
  \end{tabular}
\end{center}
\caption{Definition of numerical type $T$}
\label{Table.T}
\end{table}

\subsection{Chemical evolution}
\label{sec:Improve Chemical Evolution}
In Section~\ref{sec:Chemical Evolution}, we assumed that gas ejected from the
disk has the same metallicity as the ISM at the time that the gas was
ejected. Due to this assumption, the predicted metallicty curve for Sa and Sb
galaxies today is very high (see Figure~\ref{fig:Metallicity_Prof}, top
panel). This problem does not appear in smaller galaxies (Sc, Sd, Sm) because
the infall rate is higher than the rate of star formation, thus allowing the
accreted metal-poor gas to dilute the ISM faster than it can be enriched by
evolving stars. A comparison with observational data
\cite{pilyugin04,kewley05} leaves metal-rich outflows as the only viable
mechanism for producing the low effective yields observed in gas-rich
galaxies. Following \pcite{dalcanton06}, we parameterise the metallicity
$Z_{SN}$ of the SN ejecta as a multiple $\eta$ of the nucleosynthetic
yield. For a Salpeter IMF $\eta=~6.2-7.1$. For this paper, we adopt a value
$\eta=~4.25$ for Sa type galaxies and a smaller value $\eta=~2$ for the other
galaxy types. Furthermore, in this improved model for chemical evolution for
the accreted gas, we adopt a metallicity of $Z_{F}=~0.1{\times}$Z$_{\odot}$. The equation of
galactic chemical evolution is :
\begin{equation}
{\Sigma}_{g}dZ=~p{\psi}dt+(Z_{F}-Z)f-Z({\eta}-1){\dot{M}}_{W}
\end{equation}

\begin{figure*}	
\epsfig{file=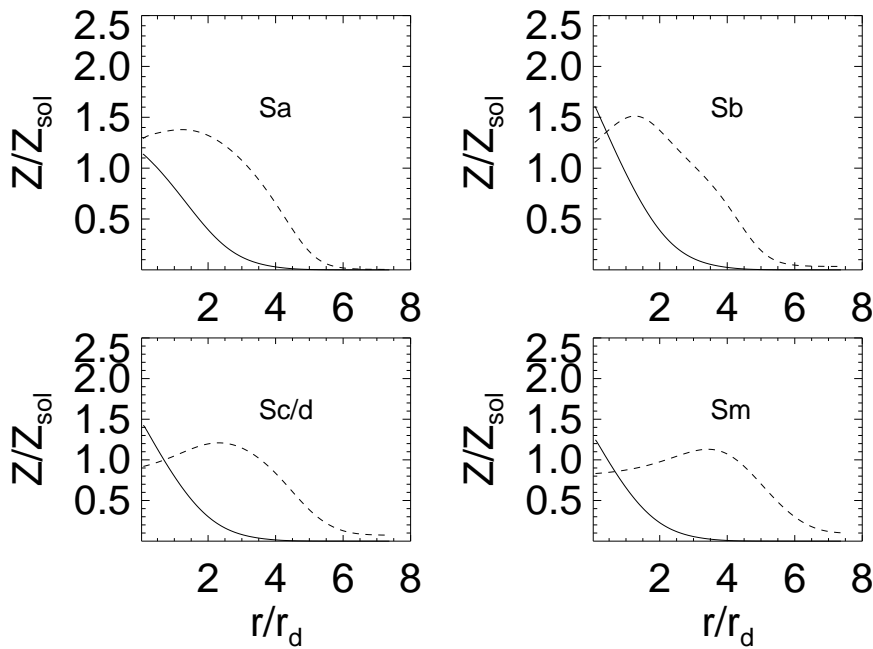,width=12cm}
\caption{The evolution of the radial distribution of metallicity for
different ages and different types of disk galaxies. The results are shown for
ages of 0.1 ({\it solid} line) and 10 Gyr ({\it dashed} line).}
\label{fig:Improve_ZModel}
\end{figure*}

In figure~\ref{fig:Improve_ZModel} we show the evolution of the radial
distribution of gas metallicity for different ages and different type of disk
galaxies. The results are shown for ages of 0.1 ({\it solid} line) and 10 Gyr
({\it dotted} line). Comparison between the model predictions and the
observational data is shown in figure~\ref{fig:int_Z}. The results come from
\pcite{kewley05} (Table 2 in this paper). For the oxygen abundance in the
solar neighbourhood, we adopt a value $12+0/H=~8.5$ \cite{pilyugin02}. The
model predictions for global metallicities are  calculated assuming that the disk
is truncated at $r/r_{d}~{\approx}~5.$

\begin{figure*}	
\epsfig{file=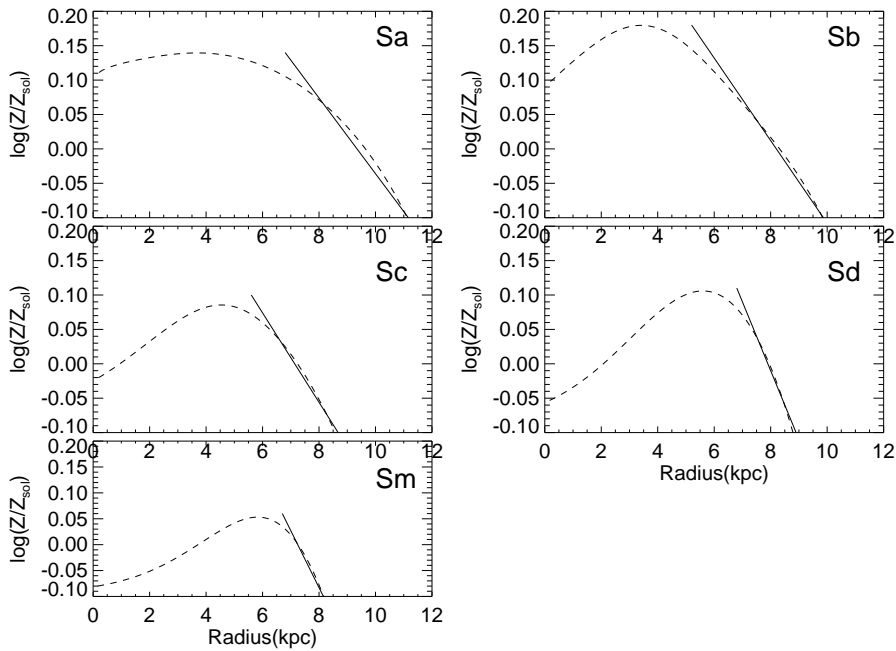,width=12cm}
\caption{Radial metallicity distribution of the gas after 10 Gyr ({\it dashed} line). The solid lines indicate metallicity slopes at the solar radius. Sa (dlog(Z)/dr= -0.055 dex kpc$^{-1}$), Sb (dlog(Z)/dr= -0.06 dex kpc$^{-1}$), Sc (dlog(Z)/dr= -0.065 dex kpc$^{-1}$), Sd (dlog(Z)/dr= -0.1 dex kpc$^{-1}$) and Sm (dlog(Z)/dr= -0.11 dex kpc$^{-1}$).}
\label{fig:Z_gradient}
\end{figure*}

\begin{figure}	
\epsfig{file=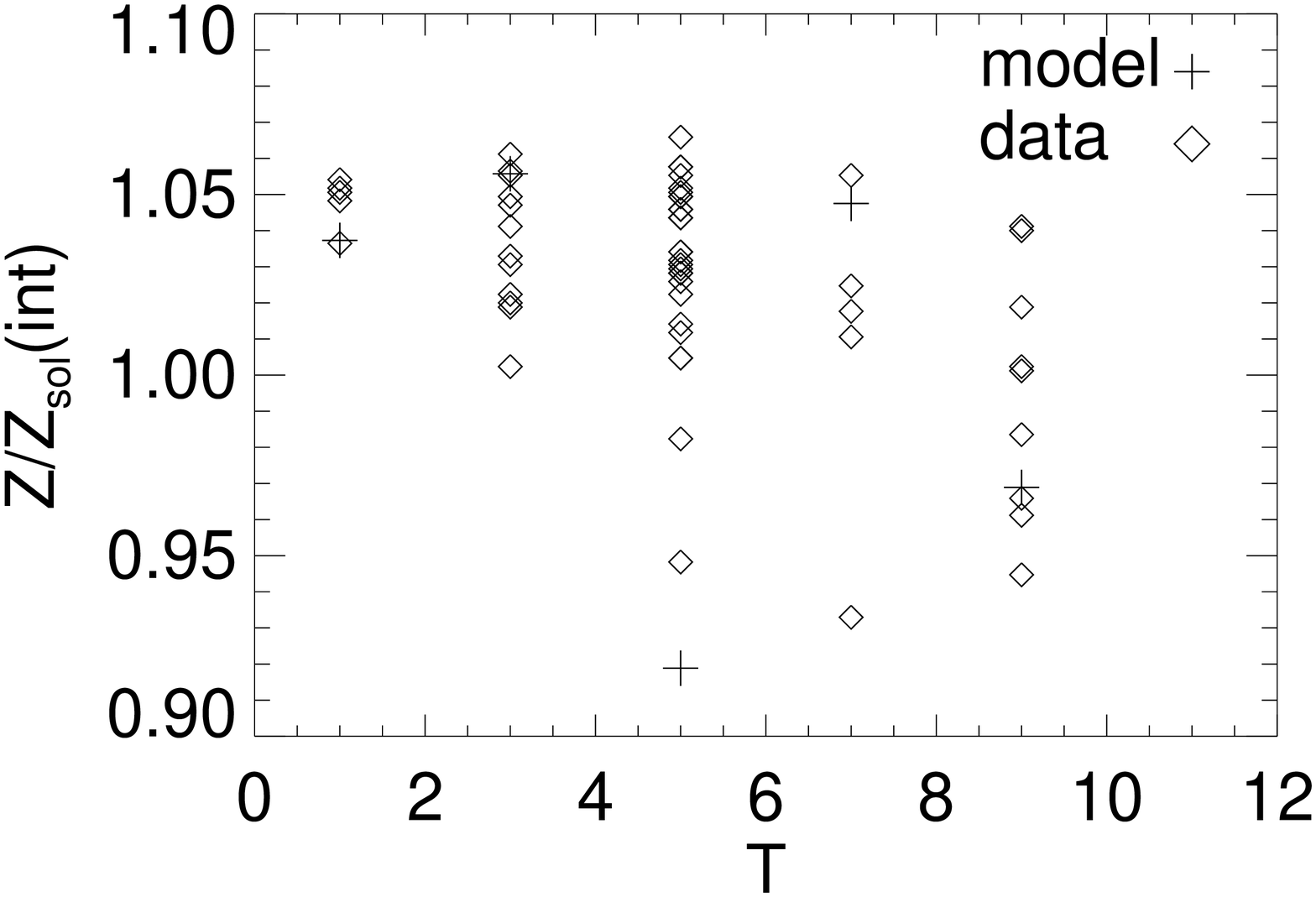,width=8cm}
\caption{The integrated metallicity is plotted versus morphological
type. Observational data({\it diamond points}) come from Kewley et al. 2005,
{\it plus points} correspond to the improved model predictions.}
\label{fig:int_Z}
\end{figure}

Figure~\ref{fig:Z_gradient} shows the radial metallicity distribution of the interstellar medium for each of our disk models after $10$ Gyr ({\it dashed} line). For the Sa model which is similar to the Milky Way the metallicity gradient is about dlog(Z)/dr=~-0.055 dexkpc$^{-1}$ at the solar radius. Observational values for the metallicity gradient of light elements X in the Milky Way are in the range of $-0.04<$dlog(X/H)/dr$<-0.08$ dexkpc$^{-1}$~\cite{chiappini01}. Our models predict that the metallicity gradient at the solar radius of the disks depends on the type, and increases as we go from early spirals to later types in agreement with observations~\cite{marquez02}. Our predictions for the metallicity gradient of later type disk galaxies are in reasonable agreement with observations~\cite{vila92} although there is a considerable scatter in the published values.

\section{Discussion}
\label{sec:discuss}

We have presented a model of global star formation incorporating supernova
feedback, gas accretion and enriched outflows in disks modelled by a
multiphase interstellar medium in a fixed gravitational potential.

A key prediction of this model is that star formation histories of different
types of disk galaxies can be explained in a simple sequence of models which
are primarily regulated by the cold gas accretion history. The distributions
of disk birth parameters presented in \pcite{kennicutt94} are reproduced
using the parameter $t_{low}$ which varies with the type of disk galaxy. Sa
galaxies are characterised by  quiescent evolution and a small value for the
$t_{low}$ parameter whereas Sb, Sc, Sd and Sm galaxies are characterised by
starbursts and relatively large values for the $t_{low}$ parameter.

A description of disk evolution in which  the onset and duration of the
gas infall history are found to be the controlling parameters is in
qualitative agreement
with standard ${\Lambda}CDM$ cosmology which predicts that protodisks reside
in dark halos with masses $\sim10^{12}\rm M_\odot $ and are in a phase of strong gas accretion with values ${>}100$ M$_{\odot}$/yr \cite{burkert06}.

The mass assembly of galaxies occurs through two main processes: hierarchical merging of smaller entities, and more diffuse gas accretion. The relative importance of the two processes cannot be easily found by cosmological simulations, since many physical parameters such as gas dissipation, star formation and feedback, are still unknown.

There are at least some twenty examples of galaxies which in HI show either signs of interactions and/or have small companions \cite{sancisi99}. This suggests that galaxies often are in an environment where material for accretion is available. Characteristic examples are the companions NGC 4565-4565A \cite{rupen91} and NGC 4027-4027A \cite{phookun92}. These companions have systematic velocities close to those of the main galaxy and HI masses less than 10{\%} of the main galaxy. The HI picture suggests the capture of a gas rich dwarf by a massive system probably to be followed by tidal disruption and accretion of the dwarf. Such examples have been seen also in the Milky Way. The discovery of the Sagittarius dwarf galaxy \cite{ibata94} shows that accretion is still taking place at the present time.

The model presented here predicts that metal-rich winds are needed
especially for Sa type galaxies in order to produce reasonable values for the
metallicity today. This prediction is in agreement with suggestions presented
in \pcite{dalcanton06} (for more detailed comments see
Section~\ref{sec:Improve Chemical Evolution}).

The major shortcoming of the present model is the failure of the chemical
enrichment model to reproduce the observed values for nuclear
metallicities. Our model predicts very high values for metallicities close to
the disk centre for Sa and Sb galaxies whereas the predicted nuclear
metallicities for Sd and Sm galaxies are quite low. This is probably due to
the high star formation rates that the model predicts in the first
$10^{8}$yr, due to our implicit assumption that the entire gas disk has
formed instantaneously at $t=~0$. The extremely high initial star formation
rate  result in highly enriched central regions, so enriched that the fresh
low metallicity accreted gas cannot dilute the gas efficiently. This is the
case especially in galaxies like Sa and Sb characterised by small infall
rates. In galaxy types like Sd and Sm, the very high infall rates in the
central regions (due to the functional form we assume for the infall model: see
Sec~\ref{sec:Infall Model}), are enough to dilute metals and result in low
values for nuclear metallicities. However the chemical evolution model is in
good agreement with observations in terms of the integrated metallicities.

In a future paper, we will  develop this model further. Along with the
improvements mentioned above, we intend to examine the effect of AGN feedback
and compare with numerical simulations.

\end{document}